\documentclass[useAMS,usenatbib]{mn2e}
\usepackage{amssymb,amsmath,latexsym,graphicx,natbib,mathrsfs}

\usepackage[usenames,dvipsnames]{color}

\title[The Nature of MSPs with Helium WD Companions]
{The Nature of Millisecond Pulsars with Helium White Dwarf Companions}

\author[S. L. Smedley, C. A. Tout, L. Ferrario and D. T. Wickramasinghe]
{Sarah L. Smedley$^1$, Christopher A. Tout$^{1,2,3}$, Lilia Ferrario$^3$\\
\newauthor
and Dayal T. Wickramasinghe$^3$\\
$^1$ Institute of Astronomy, The Observatories, Madingley Road, Cambridge, CB3 0HA\\
$^2$ Monash Centre for Astrophysics, School of Mathematics, Building 28,
Monash University, Clayton, VIC 3800, Australia\\
$^3$ Mathematical Sciences Institute, The Australian National University,
ACT 0200, Australia\\
}

\begin{document}

\date{Accepted. 2013 October 21 Received 2013 October 16; in original form 2012 December 21}
\pagerange{\pageref{firstpage}--\pageref{lastpage}} \pubyear{}

\maketitle

\label{firstpage}

\begin{abstract}
We examine the growing data set of binary millisecond pulsars that are
thought to have a helium white dwarf companion.  These systems are
believed to form when a low- to intermediate-mass companion to a
neutron star fills its Roche lobe between central hydrogen exhaustion
and core helium ignition.  We confirm that our own stellar models
reproduce a well-defined period--companion mass relation irrespective
of the details of the mass transfer process.  With magnetic braking
this relation extends to periods of less than~1\,d for a $1\,\rm
M_\odot$ giant donor.  With this and the measured binary mass
functions we calculate the orbital inclination of each system for a
given pulsar mass.  We expect these inclinations to be randomly
oriented in space.  If the masses of the pulsars were typically
$1.35\,\rm M_\odot$ then there would appear to be a distinct dearth of
high-inclination systems.  However if the pulsar masses are more
typically $1.55$ to~$1.65\,\rm M_\odot$ then the distribution of
inclinations is indeed indistinguishable from random.  If it were as
much as $1.75\,\rm M_\odot$ then there would appear to be an excess of
high-inclination systems.  Thus with the available data we can argue
that the neutron star masses in binary millisecond pulsars recycled by
mass transfer from a red giant typically lie around $1.6\,\rm M_\odot$
and that there is no preferred inclination at which these systems are
observed.  Hence there is reason to believe that pulsar beams are
either sufficiently broad or show no preferred direction relative to
the pulsar's spin axis which is aligned with the binary orbit.  This
is contrary to some previous claims, based on a subset of the data
available today, that there might be a tendency for the pulsar beams
to be perpendicular to their spin.
\end{abstract}

\begin{keywords}
stars: neutron - stars: mass-loss - stars: evolution - pulsars:
general - binaries: close
\end{keywords}

\section{Introduction}
Millisecond pulsars (MSPs) are radio pulsars with pulse periods
canonically defined to be less than $30\,\rm ms$.  About 90\,per cent
of MSPs are found in binary systems (BMSPs).  From their
spin-down rates, these are estimated to have magnetic fields of
between~$10^7$ and~$10^9\,$G, significantly lower than the $10^{12} -
10^{13}\,$G of ordinary radio pulsars \citep{Lorimer}. In the standard
rejuvenation model of BMSPs \citep{1982Alpar,1982Radhak} the field
decays and the pulsar spins up as matter is accreted from a companion
star during a previous phase of evolution as a low- or
intermediate-mass X-ray binary (LMXB/IMXB).  The discovery of several
X-ray pulsars with millisecond periods in LMXBs and of
IGR~J18245--2452, an LMXB with a rotation period of 3.9\,ms that
alternates between a rotation-powered radio source and an
accretion-powered X-ray source \citep{papitto2013} have provided
strong evidence in support of a rejuvenation model.  The precise
nature of the evolution or decay of the magnetic field during
accretion and the amount of mass accreted along the different binary
pathways remains unresolved but must depend both on the origin of the
field and the evolutionary history of BMSPs.

The lowest-mass neutron stars are expected to from by electron capture
in the cores of the most massive asymptotic giant branch stars
\citep{sugimoto1980}.  An electron degenerate core collapses, as
electrons are captured by $^{24}$Mg and~$^{20}$Ne, at a mass of
$1.375\,\rm M_\odot$ \citep{1984Nomoto}.  The cores of more
massive stars collapse only after burning to iron-group elements and
exceeding the Chandrasekhar limit.  Some of this baryonic mass,
depending on the equation of state of neutron stars,
is lost to gravitational binding energy.  \citet{2010Schwab} propose
that neutron stars formed by electron capture are typically born with
gravitational masses of around $1.25\,\rm M_\odot$ while those formed
by iron core collapse with masses around $1.35\,\rm M_\odot$.
Rejuvenation and spin up to millisecond periods requires accretion of
at least $0.1\,\rm M_\odot$ or so of material from the inner edge of a
Keplerian accretion disc.  \citet{2011Zhang}
find measured masses of the neutron stars in millisecond pulsars now to be
around $1.55\,\rm M_\odot$.  Here we are able to test whether the data
for all BMSPs with helium white dwarf companions are consistent with
such masses.

In an early study of the fastest BMSPs based on $30$ stars it was
noted that some $30\,$per cent showed the presence of a strong
interpulse separated by $180^\circ$ from the main pulse while this is
the case for only a few per cent of the general sample of radio
pulsars \citep{jayawardhana1996}. This was interpreted as evidence for
nearly orthogonal rotators, in which the magnetic dipole axis is
perpendicular to the spin axis.  Subsequently \citet*{chen1998} argued
for a bimodal distribution of both nearly orthogonal and nearly
aligned rotators among the galactic BMSPs.  \citet{ruderman1999}
presented theoretical arguments in favour of these findings based on a
specific superfluid and superconducting model for the evolution of
field in spinning accreting neutron stars.  Subsequently
\citet{1998Backer} analysed the minimum masses of a sample of low-mass
BMSPs assuming that the orbital inclinations were randomly oriented in
space but with further specific assumptions on the neutron star and
white dwarf masses and presented evidence apparently lending support
to the orthogonal rotator model.

The stellar properties of any companion along with its orbit point to
the evolutionary pathway by which the BMSP reached its current
state.  \citet{2010Hurley} made comprehensive binary population
syntheses and described the numerous routes to MSPs and the expected
relation between the final companion mass and the orbital period after
mass transfer.  Notably they included the possibility that the neutron
star began accreting as a massive ONe white dwarf that collapsed to
the pulsar after accreting sufficient mass (accretion induced
collapse).  Here we focus on what appears to be a rather common route
in which an evolved low- to intermediate-mass star fills its Roche
lobe before core helium ignition but after core hydrogen exhaustion.
This sequence is apparent in the figures of \citet{2010Hurley} whether
or not the neutron star is formed by accretion induced collapse.

In this paper we have undertaken a detailed statistical study of the
significantly enhanced current sample of millisecond pulsars with
helium white dwarf companions with their well determined mass
functions derived from quantities given in the Australia Telescope
National Facility (ATNF) pulsar catalogue.
Our study is similar to that of \citet{2005Stairs} but the increase in
data mean we can now make somewhat more significant conclusions.

Our analysis indicates that good overall agreement with the totality
of observations of BMSPs with He white dwarf companions can be
achieved if the masses of the neutron lie typically around $1.6\,\rm
M_\odot$ which is higher than what is expected from typical
electron-capture or core-collapse supernovae but is consistent with expectations of the re-cycling
hypothesis.  There is no significant difference in this consistency
between low- and high-period systems as was suggested by
\citet{2005Stairs}.  Nor can we agree with the conclusion of
\citet{1998Backer} in support of orthogonal rotators.

\section{Detailed Models of RLOF}

We compute an extensive grid of models with the Cambridge STARS code.  In
all cases we fully compute the evolution of the low-mass donor star
while the neutron star is treated as a point mass accretor.  The donors begin
their evolution as a zero-age main-sequence stars in thermal
equilibrium.  With a series of models we investigate the effect of
different binary parameters on the $M_{\rm c}$--$P_{\rm orb}$ relation
for the whole range of Case~B RLOF.  In the terminology of
\citet{1967Kippe} Case~B refers to mass transfer that begins after the
exhaustion of hydrogen fuel in a star's core but before ignition of
helium.  We refer to late Case~B when mass transfer begins
after the star is established on the giant branch, early Case~B when
it begins when the star is crossing the Hertzsprung gap and very early
Case~B when it begins at the tip of the main-sequence once the star
has a helium core but before it begins its excursion to the red in
the H--R diagram.

\subsection{Cambridge STARS Code}

We use a version of the Cambridge STARS code \citep{stars,stars2}
updated by \citet{2009Stan}.  The code features a non-Lagrangian mesh.
Convection is according to the mixing-length theory
of \citet{1958Bohm} with $\alpha_{\rm MLT}=2$ and convective
overshooting is included as described by \citet{1997Sch}.  The nuclear
species $^{1}\rm H$, $^{3}\rm He$, $^{4}\rm He$, $^{12}\rm C$,
$^{14}\rm N$, $^{16}\rm O$ and $^{20}\rm Ne$ are evolved in detail.
Opacities are from the OPAL collaboration \citep{1996Iglesias}
supplemented with molecular opacities of \citet{1994Alex}
and \citet{2005Ferg} at the lowest temperatures and
by \citet{1976Buchler} at higher temperatures.  Electron conduction is
that of \citet{1969Hub} and \citet{1970Can}.  Nuclear reaction rates
are those of \citet{1988Cau} and the NACRE
collaboration \citep{1999Angulo}.

\subsection{Period -- white dwarf mass relation}
For the donor stars the luminosity depends only on the helium core
mass and the radius depends only on the luminosity and the total mass
\citet{pacz} so there is a relation between the orbital period $P_{\rm
  orb}$ and the remnant mass $M_{\rm c}$ when the system detaches
\citep{1971Refsdal}.  We do not expect the $M_{\rm c}$--$P_{\rm orb}$
relation to depend on how the system gets to this point so that
details of how conservative is the mass transfer do not matter.  Nor
does the relation depend on the neutron star mass because for low
enough mass ratios the Roche lobe radius is proportional to the ratio
of the separation to the total mass of the system and by Kepler's
third law the cube root of the square of the orbital period is
proportional to this same ratio.  Thus $P_{\rm orb}$ is a function
only of $M_{\rm c}$.  We demonstrate that this is the case in the
following sections.
\begin{figure}
\centering 
\includegraphics[width=8cm]{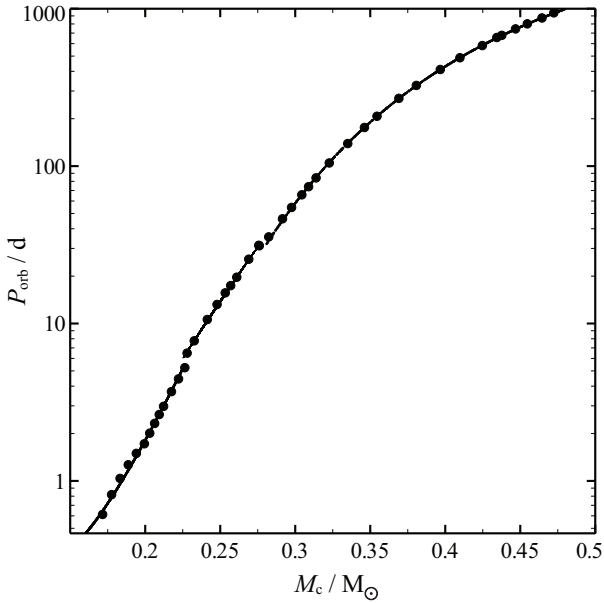}
\caption{Our Canonical $P_{\rm orb}-M_{\rm c}$ relation.  Points are a
  subset of our models made with $1\,\rm M_{\odot}$ donor stars in
  binary systems with neutron star companions of $1.55\,\rm M_{\odot}$
  (circles).  The solid line is our empirical fit
  (equation~\ref{eqfit}).}
\label{canonical} 
\end{figure}

\begin{figure}
\centering 
\includegraphics[width=8cm]{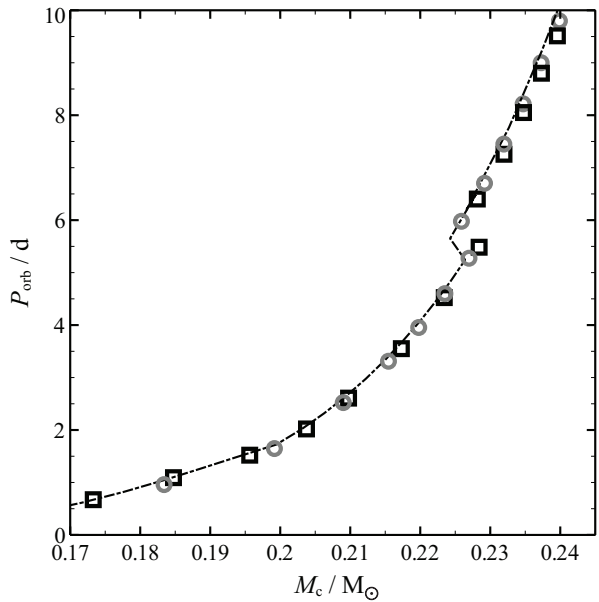} 
\caption{Variation of initial donor mass.  Models made with a
  $0.875\,\rm M_\odot$ donor are open squares and models made with a
  $1.2\,\rm M_\odot$ donor are light open circles.  The dot-dashed
  line is our canonical relation from $1\,\rm M_\odot$ donors.}
\label{comparerelations} 
\end{figure}

Though the mass transfer to MSPs is likely to be non-conservative we
first consider the canonical, perhaps extreme, fully conservative case
in which all of the envelope of the donor star is transferred to and
accreted by its companion and for which total angular momentum is
conserved.  We model initial donor masses of 0.875, 1 and~$1.2\,\rm
M_\odot$.  We give the companion an initial mass of $1.55\,\rm
M_\odot$.  We demonstrate later that the relation is not affected by
the non-conservative nature of the mass transfer.  For each donor mass
we evolve a set of 300~models with initial periods to cover the full
range of Case~B RLOF for each of the different donor masses.  Magnetic
braking, as discussed in more detail in the next section, is necessary
to reach the lowest periods and core masses.  Fig.~\ref{canonical}
shows a subset of the $1\,\rm M_{\odot}$ donor models plotted with an
empirical fit for the relation (section~\ref{secfit}).  In
Fig.~\ref{comparerelations} we plot the $P_{\rm orb}-M_{\rm c}$
relation for a selection of models with each of the three initial
donor masses.  There is no perceptible difference.  Lower-mass donor
stars exhibit very similar behaviour to the $1\,\rm M_\odot$ donor
once they evolve past the main sequence.  However they are not
expected to have evolved within the lifetime of the Galaxy.  Higher
mass donors, once established on the giant branch undergo unstable
RLOF which probably leads to common-envelope evolution and perhaps
merging of the stars.  Thus in general we do not expect the donor star
mass to lie outside the range $0.875-1.2\,\rm M_\odot$ very often and
so we take the $P_{\rm orb}-M_{\rm c}$ relation derived from $1\,\rm
M_\odot$ donors to be canonical.  There are some discontinuities
apparent for all donor masses around a remnant mass of $0.225\,\rm
M_\odot$ caused by dredge up of the previously burnt stellar interior
by the deepening convective envelope.

\subsection{Short Periods and Magnetic Braking}
The minimum white dwarf mass
that can be left by a given donor depends on the
Sch\"onberg--Chandrasekhar mass \citep{schonberg1942} of its core.
Beyond this the core cannot remain isothermal and still support the
stellar envelope.  After this the star evolves rapidly across the HG.
If evolved fully conserving mass and angular momentum our
$1\,\rm M_\odot$ donor leaves a minimum remnant mass of $0.22\,\rm
M_\odot$.

A group of very short orbital period MSPs are thought to have evolved
from LMXBs \citep{fabian1983} for which the mass transfer is driven by orbital angular
momentum loss rather than stellar evolution.  In these
systems the, typically low initial mass ($M_{\rm d} < 0.9\,\rm M_\odot$)
donor star does not develop a degenerate core and final
periods are a few hours.

With their shorter hydrogen-burning lifetime intermediate-mass donors, those around
$1\,\rm M_\odot$ and more, can develop helium cores but are also expected to suffer magnetic braking if their
orbital period is short enough.  If such an evolved intermediate-mass star begins RLOF before this
(very early Case~B) its limiting core mass is reduced as it transfers
mass and so consequently behaves as if it were a star with a somewhat
lower initial total mass.  There is a bifurcation period below which systems evolve as
LMXBs to very low $P_{\rm orb}$ and above which they evolve to larger
$P_{\rm orb}$ transferring mass as red giants or subgiants.  We are
interested in these latter systems which end up detached with $P_{\rm
  orb} \ga 1\,$d.

We include magnetic braking at the empirical rate of \citet{1981Verbunt},
\begin{equation}
\frac{\dot{J}}{J} = -0.5 \times 10^{-28}\,{\rm s^2\,cm^{-2}} f_{\rm mb}^{-2} \frac{IR_{\rm d}^{2}}{a^{5}} \frac{GM^{2}}{M_{\rm a}M_{\rm d}}\,\rm s^{-1},
\end{equation}
where $R_{\rm d}$ is the radius of the donor star, $I$ its moment of
inertia, $a$ the semi-major axis of the orbit, $M$ the total mass of the binary system,
$M_{\rm a}$ the mass of the accretor and $M_{\rm d}$ is the mass of
the donor.  The factor $f_{\rm mb}$ was chosen to fit the equatorial
velocities of G~and K~type stars \citep{smith1979}.

\begin{figure}
\centering 
\includegraphics[width=8cm]{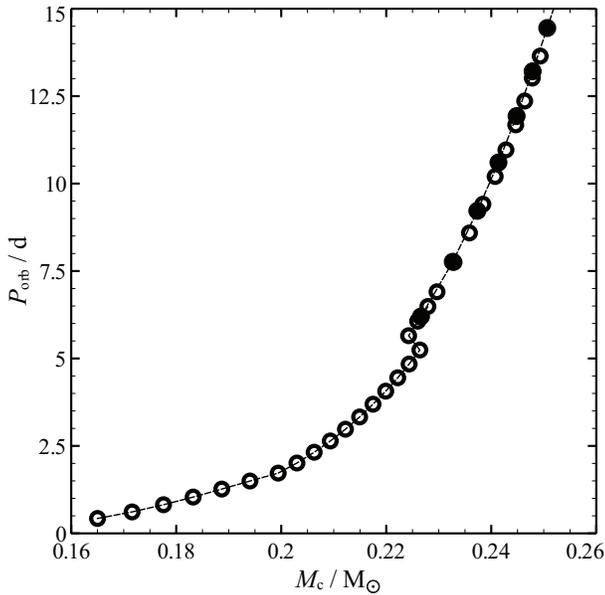} 
\caption{Detail of the period--core mass relation for critically braked systems
(open circles) compared with non-braked systems (solid
circles).  All models have $1.55\,\rm M_{\odot}$ 
neutron star that accreted from an initially $1\,\rm M_{\odot}$ donor.}
\label{mb} 
\end{figure}

Fig.~\ref{mb} shows detail of the low-period end of the $P_{\rm
  orb}-M_{\rm c}$ relation for systems evolved with and without
magnetic braking.  The form of the magnetic-braking law is such that
the initial orbital periods for which it can produce lower-period
systems with He white dwarfs is finely balanced between 2 and~2.2\,d to produce
final periods between 0.5 and~6\,d.

\subsection{Examples of evolution} 

\begin{figure}
\centering 
\includegraphics[width=8cm]{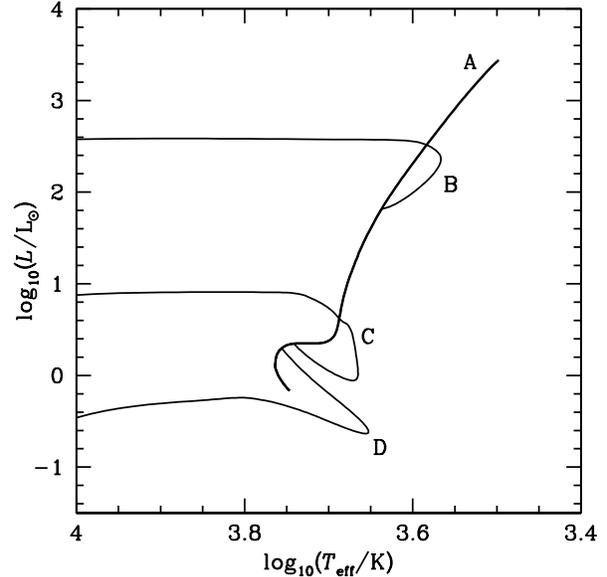} 
\caption{A Hertzsprung-Russell diagram, in the bolometric
luminosity $L$--effective temperature $T_{\rm eff}$ plane, for the evolution of an
isolated $1\,\rm M_{\odot}$ star (labelled~A), along with the
evolution of similar stars in binary systems with neutron star
companions of $1.55\,\rm M_{\odot}$.  All have active magnetic braking.  
Model~B has an initial orbital
period of $2\,$d, model~C $2.14\,$d and~D $20\,$d.}
\label{hrdsample} 
\end{figure}

Fig.~\ref{hrdsample} shows a Hertzsprung--Russell diagram for three binary star
models (B, C and~D) that begin RLOF at different stages along with a
track~(A) for single star evolution.  Model~B begins RLOF only once
established as a red giant, late Case~B.  Model~C is crossing the
Hertzsprung gap, begins RLOF as a subgiant, early case~B, while
model~D begins mass transfer with an isothermal gas-pressure supported
core, very early Case~B.
By the time the stars in models C and~D detach they have
established degenerate He cores and are ascending the red giant
branch.  The larger the initial period, the further up the giant
branch the RLOF begins.  Mass transfer strips the envelope to reveal
the stars' cores which then cool to become white dwarfs.

\subsection{Empirical fit}
\label{secfit}

\par A good empirical fit to $P_{\rm orb}-M_{\rm c}$ relation for all
the models with $1\,\rm M_\odot$ donors, including those with magnetic
braking, is given by 
\begin{equation}
\label{eqfit}
{P_{\rm orb}}/{\rm d} = \exp(a+(b/(M_{\rm c}/\rm M_{\odot}))+c(\ln(\it M_{\rm c}/\rm M_{\odot}))),
\end{equation}
 with
\begin{equation}
 (a,b,c)  = 
  \begin{cases}
    (22.902, 3.097, 23.483) & \rm for\,\, 0.17 \le {\it M}_{\rm c} < 0.225\\
    (14.443, 0.335, 9.481) & \rm for\,\, 0.225 \le {\it M}_{\rm c} < 0.28\\
    (11.842, \mbox{-}3.831, \mbox{-}4.146) & \rm for\,\, 0.28 \le {\it M}_{\rm c} < 0.48.\\
  \end{cases}
\end{equation}

\subsection{Measured pulsar masses}

\begin{table*}
\centering
\caption{BMSPs with measured pulsar and companion masses.  Two
  systems with $P_{\rm spin} > 30\,$ms are included for completeness
  but are excluded from our analysis.\label{pulsartable1}}\vspace{5pt}
\begin{tabular}{| l | r | r | l | l | r ||}
\hline
Name & $P_{\rm spin} /\rm ms$ & $P_{\rm orb}/\rm d$ & $M_{\rm p}/\rm M_{\odot}$ & $M_{\rm comp}/\rm M_{\odot}$ & Reference \\
\hline
J0348+0432  & 39.12 &    0.103\!\!\! & $2.01 \pm 0.04$ & $0.172  \pm 0.003$  & \cite{antoniadis2013} \\
J0751+1807  &  3.48 &    0.26        & $1.26 \pm 0.14$ & $0.191  \pm 0.015$  & \cite{2008Nice}       \\
J1738+0333  &  5.85 &    0.35        & $1.47 \pm 0.06$ & $0.181  \pm 0.006$  & \cite{2012Antoniadis} \\
B1957+20    &  1.61 &    0.38        & $2.40 \pm 0.12$ & $0.035  \pm 0.002$  & \cite{2011Gonzalez}   \\
J1012+5307  &  5.26 &    0.60        & $1.6  \pm 0.2$  & $0.16   \pm 0.02$   & \cite{2005vankerk}    \\
J1910--5959A&  3.27 &    0.84        & $1.33 \pm 0.11$ & $0.180  \pm 0.018$  & \cite{2012Corongiu}   \\
J1909--3744 &  2.95 &    1.53        & $1.438\pm 0.024$& $0.2038 \pm 0.0022$ & \cite{2005Jacoby}     \\
J0437--4715 &  5.76 &    5.74        & $1.76 \pm 0.02$ & $0.254  \pm 0.014$  & \cite{2008Verbiest}   \\
J1614--2230 &  3.15 &    8.69        & $1.97 \pm 0.04$ & $0.500  \pm 0.006$  & \cite{demorest2010}   \\
B1855+09    &  5.36 &   12.33        & $1.6  \pm 0.2 $ & $0.270  \pm 0.025 $ & \cite{2004Splaver}    \\
J1910+1256  &  4.99 &   58.47        & $1.6  \pm 0.6$  & $0.30 - 0.33$       & \cite{2011Gonzalez}   \\
J1713+0747  &  7.99 &   67.83        & $1.3  \pm 0.2$  & $0.28   \pm 0.03$   & \cite{2005Splaver}    \\
J1853+1303  &  4.09 &  115.65        & $1.4  \pm 0.7$  & $0.33 - 0.37$       & \cite{2011Gonzalez}   \\
J2016+1948  & 64.95 &  635.04        & $1.0 \pm 0.5$   & $0.43 - 0.47$       & \cite{2011Gonzalez}   \\
\hline  
\end{tabular}
\end{table*}

\begin{figure}
\centering 
\includegraphics[width=8cm]{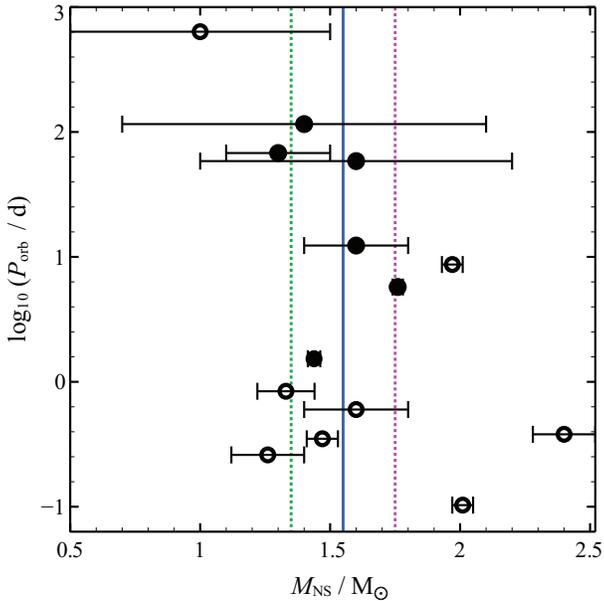} 
\caption{Neutron star masses and orbital periods of
the observed BMSPs listed in Table~\ref{pulsartable1}.
Solid circles are the systems that fit our selection criteria while
open circles are those that are excluded for various reasons.  
Vertical lines indicate masses
of 1.35, 1.55 and~$1.75\,\rm M_\odot$.}
\label{nsmass} 
\end{figure}

Table~\ref{pulsartable1} lists observed MSP -- white dwarf binary systems that have
measured pulsar masses and companion masses.  These masses are
generally higher than those found for isolated pulsars
\citep{2011Zhang}.  This may be due to either a range in initial
neutron star mass or accretion during the mass transfer.  \citet{theo}
noticed a negative correlation between the orbital period and the mass
of the pulsar.  In Fig.~\ref{nsmass} we plot the data for the systems
in Table~\ref{pulsartable1}.  Any correlation is actually rather
weak.
Though we list all the measured masses here we note that all of those
with orbital periods of less than $1\,$d are excluded from our
analysis because they may have undergone Case~A mass transfer, while the
star is still burning hydrogen in its core, rather than
Case~B mass transfer.  Indeed B1957+20 is a black widow pulsar
\citep{1988Fruchter2} with a low-mass main-sequence star, rather than a
white dwarf.  J1910--5959A is also excluded
because it is a member of a globular cluster \citep{2012Corongiu}. J2016+1948 is
 excluded because it has $P_{\rm spin} > 30\,\rm ms$ and J1614--2230 is excluded 
because it has a companion mass larger than $0.472\,\rm M_\odot$.
  
\begin{figure}
\centering 
\includegraphics[width=8cm]{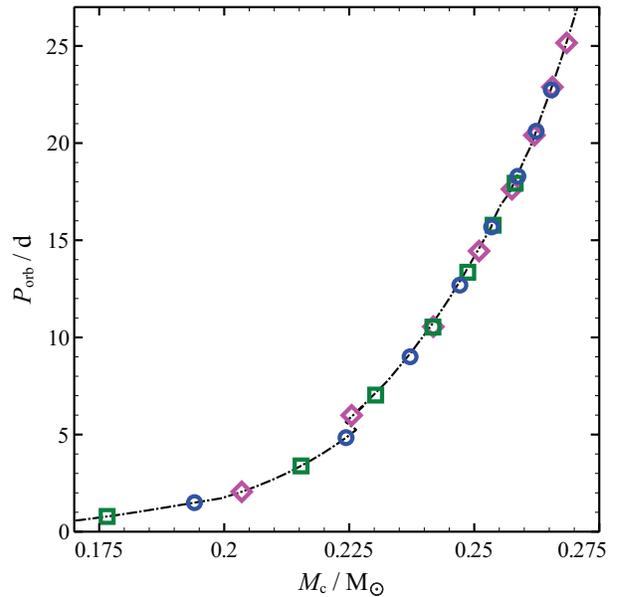} 
\caption{The effect of neutron star mass on the
$M_{\rm c}$--$P_{\rm orb}$ relation.  Three cases are shown here, all
models have a $1\,\rm M_{\odot}$ donor star and were computed with
initial periods from
$2\,$d to~$2.4\,$d.  Open squares have
an initial neutron star mass of $1.35\,\rm M_{\odot}$, open circles
$1.55\,\rm M_{\odot}$ and open diamonds $1.75\,\rm M_{\odot}$.  The
dot-dashed line is our empirical fit.  As expected the neutron star mass has no
effect on the $P_{\rm orb}-M_{\rm c}$ relation.}
\label{changemass} 
\end{figure}

\cite{2010DeVito} looked at the effect of the neutron star mass on
these systems and found that, for systems of given initial orbital
period and initial donor star mass, the mass of the neutron star
heavily affects the evolution. However they conclude that the
$M_{\rm c}$--$P_{\rm orb}$ relation is insensitive to the initial
neutron star mass, confirming the prediction made by \cite{1995Rappa}.
We tested the effect of the neutron star mass on the evolution of our
binary systems by computing models that initially had a $ 1\,\rm M_{\odot} $
donor star and neutron star masses of $1.35$, $1.55$ and~$1.75\,\rm
M_{\odot}$.  We made 100 models for each case with initial orbital
periods spanning the entire $M_{\rm c}$--$P_{\rm orb}$ relation.  
Fig.~\ref{changemass} shows the $M_{\rm
  c}$--$P_{\rm orb}$ relations for a subset of these models with 
initial periods between 2 to 2.4\,d.  The $M_{\rm c}$--$P_{\rm orb}$
relation itself is not affected by the neutron star mass but, for a
given initial orbital period, the higher the neutron star mass, the
more massive the predicted white dwarf remnant.  The higher mass of
neutron star lies further up the relation but not off it.

However the neutron star mass does matter when we compute orbital
inclinations (equation~\ref{eqinc}).  For now we conclude that masses of
MSPs in systems that could have helium white dwarf companions lie
between about 1.35 and $1.75\,\rm M_\odot$ and cluster around
$1.55\,\rm M_\odot$. 

\subsection{RLOF Efficiency}

Relaxing conservative evolution further, we allow mass to be lost
from either star during mass transfer.  The total
angular momentum $J$ is the sum of the orbital angular momentum and
the rotational (or spin) angular momentum of each star.  The latter is
typically only $1-2\,$per cent of the orbital angular momentum and so
can usually be ignored though we do include it in our detailed stellar
models.  \cite{theo} argued that the accretion process is inefficient,
with only a small portion of transferred matter accreted by the neutron
star.  The rest is ejected.  \cite{spinup} confirmed that the
accretion process in these systems is non-conservative.  Only
0.1\,M$_{\odot}$ of material from the inner edge of the disc is needed
to spin up the neutron star to periods below one millisecond.
\cite{2012DeVito} computed the evolution of a set of models with
initial donor star between $0.5$ and~$3.5\,\rm M_{\odot}$, initial
orbital periods between 0.175 and 12\,d and initial neutron stars
masses between $0.8$ and~$1.4\,\rm M_{\odot}$, with varying degrees of
mass transfer efficiency.  Following their approach we define
efficiency parameters $\alpha$, the fraction of mass lost by the donor
that is transferred to the accretor and $\beta$, the amount of matter
actually captured by the accretor.  A fraction $1-\alpha$ of the mass
lost by the donor then carries off the specific angular momentum of
the donor while a fraction $\alpha(1-\beta)$ of the mass lost by
the donor carries off the specific angular momentum of the accretor.  So
\begin{equation}
\dot{M_{\rm a}} = -\alpha\beta\dot{M_{\rm d}},
\end{equation}
and
\begin{equation}
\dot{J} = (1-\alpha)\dot{M_{\rm d}} a_{2}^{2}
\omega + \alpha (1-\beta) \dot{M_{\rm d}} a_{1}^{2} \omega.
\end{equation}
Thence
\begin{equation}
\frac{\dot{J}}{J} = \frac{(1-\alpha)\dot{M_{\rm d}} a_{2}^{2} \omega +
  \alpha (1-\beta) \dot{M_{\rm d}} a_{1}^{2} \omega}{(M_{\rm a}M_{\rm
    d}/(M_{\rm a} + M_{\rm d}))a^{2} \omega},
\end{equation}
where $J = M_{\rm a}M_{\rm b}a^2\omega/(M_{\rm a} + M_{\rm b})$ is the orbital angular
momentum for semi-major axis $a$ and orbital angular velocity $\omega
= 2\pi/P_{\rm orb}$.
This simplifies to
\begin{equation}
\frac{\dot{J}}{J} = \frac{\dot{M_{\rm d}}}{M_{\rm d}} \left[ \frac{1- \alpha + \alpha (1-\beta) q^{2}}{1-q} \right].
\end{equation}
where the mass ratio $q = M_{\rm d}/M_{\rm a}$.

\begin{figure}
\centering 
\includegraphics[width=8cm]{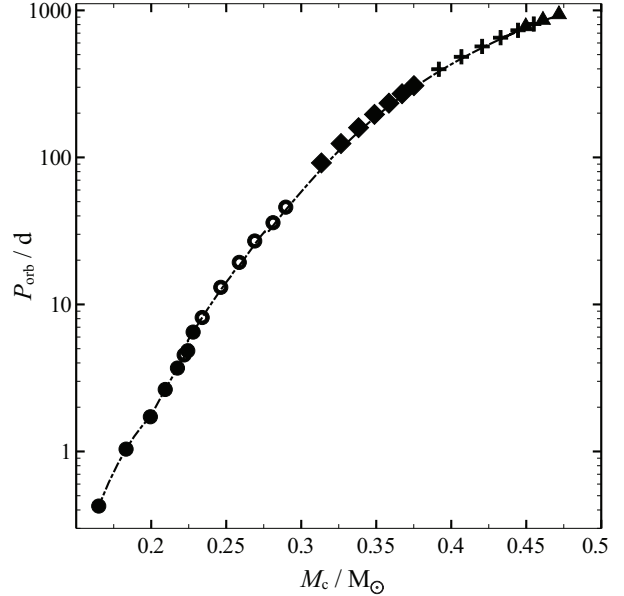} 
\caption{The effect of varying the RLOF efficiency on the $M_{\rm
    c}$--$P_{\rm orb}$ relation.  Five cases are shown, 1)~$\alpha =
  \beta = 1$ (solid circles), 2)~$\alpha = \beta = 0$ (open circles),
  3)~$\alpha = 1$ and $\beta = 0.6$ (diamonds), 4)~$\alpha = 0.6$ and
  $\beta = 0$ (crosses) and 5)~$\alpha = 1$ and $\beta = 0$
  (triangles).  Each set initially has $M_{\rm d} = 1\,\rm M_\odot$,
  $M_{\rm a} = 1.55\,\rm M_\odot$, and $2 < P_{\rm orb}/{\rm d} <
  2.19\,$.  The dot-dashed line is our canonical relation.}
\label{changeRLOF} 
\end{figure}

We evolved sets of 100~models for combinations of $\alpha$
and~$\beta$, with initial orbital periods in the range 2 to 3\,d.
Fig.~\ref{changeRLOF} illustrates the effects on
the $M_{\rm c}$--$P_{\rm orb}$ relation.  As expected the relation
between $M_{\rm c}$ and $P_{\rm orb}$ is not much affected by the
non-conservative nature of the RLOF but, for a given initial orbital
period, increasing $\alpha$ or decreasing $\beta$ leaves a more massive core and
consequently a longer final period because, even though the escaping
material carries away angular momentum, the total system mass falls.

\subsection{Comparison of Detailed Models and Data}

\begin{figure}
\centering 
\includegraphics[width=8cm]{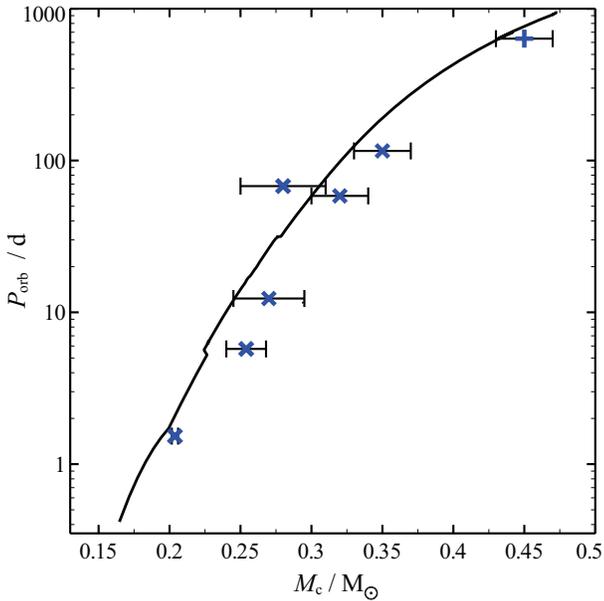} 
\caption{MSPs with measured companion masses in the $M_{\rm
    c}$--$P_{\rm orb}$ plane. The solid line is our canonical relation.
  We do not include systems excluded from our analysis except for the
  635\,d system J2016--2230 (shown by $+$).  It appears to lie precisely
 on the relation but has a spin period of $65\,$ms.}
\label{modplusdatclose} 
\end{figure}

The systems with measured neutron star masses listed in
Table~\ref{pulsartable1} also have measured companion masses.  We again
ignore systems with $P_{\rm orb} < 1\,$d.  Note that
J1614--2230 is also excluded because to fit the data and the $P_{\rm
  orb}-M_{\rm c}$ relation would require too small a neutron star
mass (see Section~\ref{orbinc}).  This is encouraging because it is indeed found to have a
companion of $0.5\,\rm M_\odot$, too massive for a He white dwarf.  We
plot the remaining data in Fig.~\ref{modplusdatclose} and find that the fit with
our canonical relation is good given the measurement errors.  Note that 
J2016+1948 is also formally excluded because its spin period exceeds
30\,ms but we plot it here because it lies on the canonical $P_{\rm
  orb}-M_{\rm c}$ relation.  This suggests that it has passed through
a similar evolution but has just not been spun up so much.

\subsection{Donor Star Metallicity}

\begin{figure}
\centering 
\includegraphics[width=8cm]{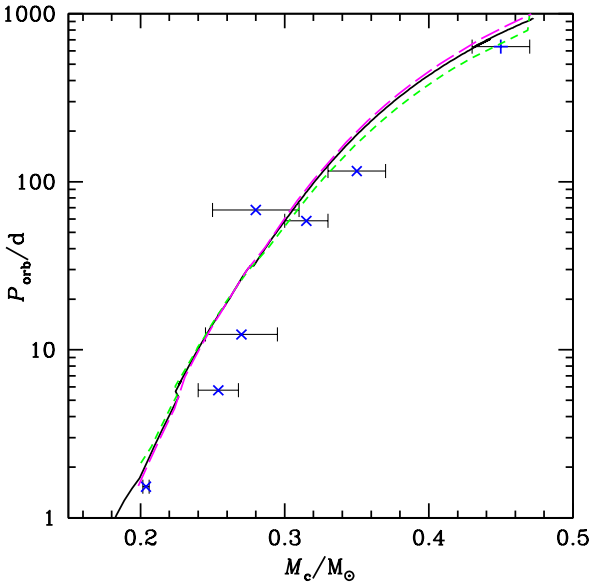} 
\caption{The effect of donor star metallicity on the $M_{\rm
    c}$--$P_{\rm orb}$ relation.  Models are all for accretion on to a
  $1.55\,\rm M_{\odot}$ neutron star from a $1\,\rm M_{\odot}$ donor.
  The solid line has $Z = 0.02$, the
  the short-dashed line $Z=0.01$ and the long-dashed line $Z=0.03$.  The effect of changing
  metallicity over the range that might be expected in the Galactic field is
  still rather small but larger than any effect owing to the
  nature of the mass transfer.}
\label{metals} 
\end{figure}

We are concentrating on BMSPs in the Galactic field and so do not
expect the metallicity to vary much from solar.  However
we investigated how the metallicity of the donor star affects our
$M_{\rm c}-P_{\rm orb}$ relation by evolving two further sets of
models with $Z = 0.01$, half solar, and $Z = 0.03$, one and a half solar. 
As metallicity falls opacity drops and stars are smaller for the same 
luminosity and hence core mass so the final orbital period for a given 
white dwarf mass is smaller. The
results are shown in Fig.~\ref{metals}.  The change in the relation with 
metallicity is larger than for any of the other effects we have tested but all three models
remain in good agreement with the observed systems.

\subsection{Comparison with Previous Work}
The relation between the mass of the white dwarf remnant and the
orbital period after mass transfer ($M_{\rm c}$--$P_{\rm orb}$) has
been computed a number of times in the past.  \citet{1995Rappa} used
11~stellar models with various masses between $0.8$ and~$2\,\rm
M_{\odot} $, computed by \citet{1994Han} with Eggleton's STARS code
\citep{stars} to obtain the relation between the radius of a star on
the red giant branch and its core mass in the range $0.15$
to~$1.15\,\rm M_{\odot}$.  They then used this to derive a $M_{\rm
  c}$--$P_{\rm orb}$ relation.  \citet{theo} conducted a detailed
study of the thermal response of the donor star to mass loss in order
to examine the evolution of the mass transfer (non-conservative in
this case) and to monitor its stability.  They used Eggleton's STARS
code to compute 121~models of the evolution of different binary
systems, with a low-mass donor and a neutron star companion, through
stable mass transfer.  Their donor masses between $1$
and~$2\,\rm M_{\odot}$, a neutron star of $1.3\,\rm M_{\odot}$ and
initial orbital periods between 2 and~$800\,$d.  They obtained $M_{\rm
  c}$--$P_{\rm orb}$ relations and devised fitting formulae for
various chemical compositions.  \citet*{2002Pod} made a systematic
study of low- and intermediate-mass binary systems with donor masses
from $0.6$ to~$7\,\rm M_{\odot}$ \citep[a much wider range
  than][]{theo} a neutron star of $1.4\,\rm M_{\odot}$ and initial
orbital periods between 0.17 and~$100\,$d.  They used an updated
version of the stellar evolution code described by \citet{1967Kipper}
to make 100~binary star models, in which half of the mass lost by the
donor was accreted by the companion.  Their models showed an enormous
variety of evolutionary channels which they suggest may be the reason
behind the diversity in the observed population of these systems.
\citet{2010DeVito} made many detailed evolutionary models with a wide
range of donor ($0.5$ to~$3.5\,\rm M_{\odot}$) and accretor ($0.8$
to~$1.4\,\rm M_{\odot}$) masses and orbital periods at the onset of
Roche lobe overflow from 0.5 to~$12\,$d with a code described by
\citet{2003Ben}.  Their models show signs of departure from the
Rappaport $M_{\rm c}$--$P_{\rm orb}$ relation at core masses below
$0.25\,\rm M_\odot$ but are in better agreement with the $M_{\rm
  c}$--$P_{\rm orb}$ relation found by \citet{theo}.  They studied the
dependence of the neutron star mass on the $M_{\rm c}$--$P_{\rm orb}$
relation and later the dependence of the mass transfer efficiency on
the $M_{\rm c}$--$P_{\rm orb}$ relation \citep{2012DeVito}.
\cite{2012Shao} explored low- and intermediate-mass systems, similarly
to \cite{2002Pod}, with the Cambridge STARS code, with donor masses of
$1$ to~$6\,\rm M_{\odot}$ and accretor masses of $1.0$ to~$1.8\,\rm
M_{\odot}$.  They modelled the $M_{\rm c}$--$P_{\rm orb}$ relation for
donors from $1$ to~$5\,\rm M_{\odot}$, compared their relations to
observed possible MSP--helium white binaries and explored magnetic
braking in these systems.
\begin{figure}
\centering 
\includegraphics[width=8cm]{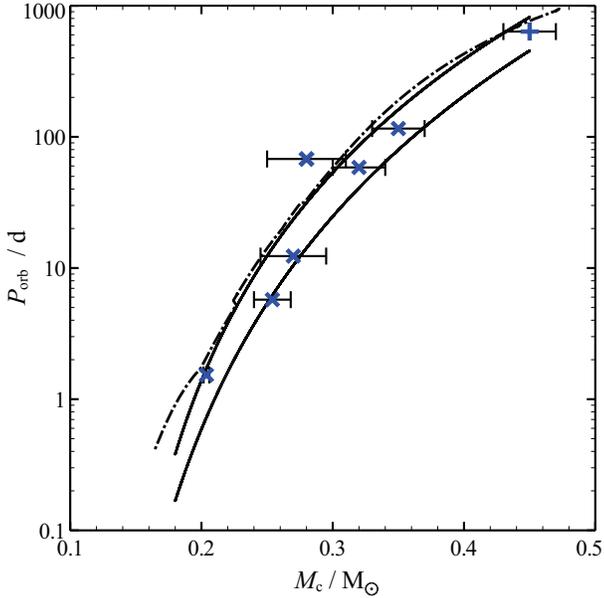} 
\caption{Comparison with the work of Tauris \& Savonije (1999). The dot-dashed line is our canonical relation. The two
solid curves are fits of Tauris \& Savonije (1999) for population I
stars (leftmost) and population II stars (rightmost).}
\label{tauris} 
\end{figure}

The fits of \citet{theo} have been most extensively used in the past so we compare our models with theirs in Fig.~\ref{tauris}.  Their fits are of the form
\begin{equation}
\frac{M_{\rm wd}}{\rm M_{\odot}} = \left( \frac{P_{\rm orb}}{b}\right)^{1/a} + c,
\end{equation}
with
\begin{equation}
 (a,b,c)  = 
  \begin{cases}
    (4.50, \quad 1.2 \times 10^{5}, \quad 0.120) & \rm Pop.I,\\
    (5.00, \quad 1.0 \times 10^{5}, \quad 0.100) & \rm Pop.II,\\
  \end{cases}
\end{equation}
for population~I stars and population~II stars.  
Our models compare well with their population~I stars.  Despite the
fact that lower metallicities are a better fit to two of the
measured companion masses we use solar metallicity because we exclude
BMSPs in globular clusters. The remaining systems fit better or equally well 
at solar metallicity.

\section{Selection of BMSPs}

\begin{figure}
\centering
\includegraphics[width=8cm]{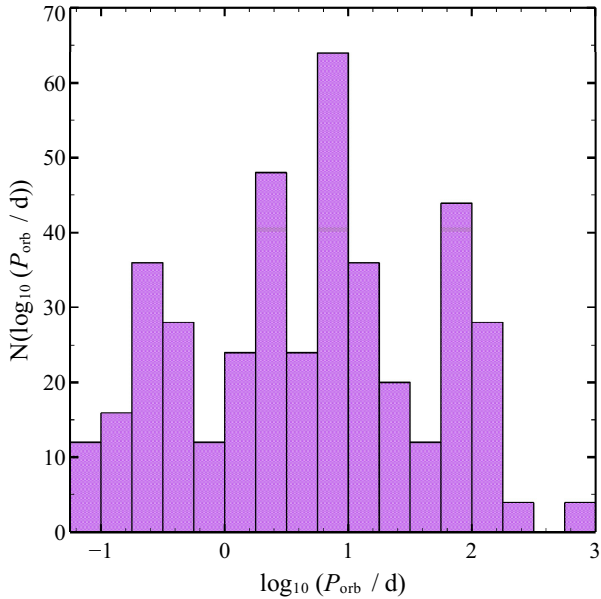}
\caption{The distribution of all BMSP orbital
  periods less than 1,000\,d.  Those systems with $P_{\rm orb} > 940\,$d are not expected to
  have helium white dwarf companions.  Below the gap apparent around $P_{\rm
    orb} = 1\,$d we can expect contamination from systems that have
  undergone Case~A mass transfer and now have low-mass hydrogen-rich
  companions.
\label{obsmspperiods}}
\end{figure}

The ATNF catalogue \citep{2005Man} currently lists 293~BMSPs
with $P_{\rm spin} \leq 30\,\rm ms$.  Of these 101 are in the Galactic
field and have measured binary mass functions.  It is important to
exclude all BMSPs that have not been recycled by accretion from a red
giant with a helium core but otherwise we wish to test the hypothesis that all
remaining systems did indeed form by this route.  We exclude any BMSPs in globular
clusters because their formation can be more convoluted
\citep{Lorimer}.
B1257+12 can be excluded directly because the orbit is for its 
planet of mass only $0.02\,\rm M_\oplus$ \citep{konacki2003} and J1502--6752 
is excluded because it
has an ultra-light companion \citep{2005Man}.  Nor did we include B1620--26
because it is identified as being a multiple system in the ATNF pulsar 
catalogue so the other components in the system
could have altered its evolution and J1903+0327 because it has a main-sequence 
star companion and is in a multiple system.
Stars below $2.3\,\rm M_\odot$
suffer a helium flash when their degenerate core reaches $0.472\,\rm
M_\odot$.  This is the maximum mass that can be reached by a helium
white dwarf companion and corresponds to a final orbital period of
940\,d.  Stars shrink during core helium burning and by the time they
have grown sufficiently to fill their Roche lobes again their cores
are generally above $0.5\,\rm M_\odot$.  There is therefore a
distinction between MSPs recycled by red giants with helium cores
($P_{\rm orb} < 940\,$d) and those recycled by asymptotic giants
($P_{\rm orb} > 940\,$d).  This distinction is apparent in
the figures of \citet{2010Hurley}.  We therefore impose an upper
period limit of 940\,d.  At low orbital periods we must exclude
systems in which mass transfer began while the donor star was still on
the main sequence (Case~A).  The observed orbital period distribution
for BMSPs appears to have a gap between $0.73\,$d and $1.19\,$d
(Fig.~\ref{obsmspperiods}).  Above this we can easily form systems
with helium white dwarf companions when magnetic braking is included
in our models.  Below it there is likely to be contamination by
systems that have undergone Case~A mass transfer and still have a
hydrogen-rich companion.  We assume that the gap represents a good
discriminator between the two populations and so exclude all systems
with $P_{\rm orb} < 1\,$d.  All the remaining systems are listed in
Table~\ref{mspsfull} which can be found in the appendix.  In practice we
might expect most of the MSPs to have formed from relatively low-mass
progenitors in electron-capture supernovae.  In such stars degenerate
cores collapse at $1.375\,\rm M_\odot$ \citep{1984Nomoto}.  They may
lose about $0.1\,\rm M_\odot$ of their gravitational mass depending on
the the equation of state of the neutron star but then probably need
to accrete a further $0.1\,\rm M_\odot$ to be spun up to millisecond
periods.  The measured masses listed in Table~\ref{pulsartable1}
suggest that the MSPs usually accrete more than this.  Nevertheless, in the next
section we describe how we further exclude systems that cannot fit the model with a
neutron star less massive than $1.2\,\rm M_\odot$ and we use this rather
conservative minimum to avoid excluding systems with the lowest
measured masses of pulsars \citep{2011Zhang}.  Only three systems need
have a pulsar mass $1.2 < M_{\rm a}/{\rm M_\odot} < 1.35$.

\section{Orbital Inclinations}

\label{orbinc}

In the recycling model the MSP has been spun up by accretion through a
disc of material transferred from a binary companion tidally locked to
the binary orbit.  The spin of the pulsar should then be aligned with
the orbit.  We expect orbital inclinations $i$ to the line of sight to
be uniformly distributed over a sphere so that the distribution of
$\cos i$ should be flat. 

\cite{1998Backer} explored the spin periods, ages and magnetic fields
of Galactic helium white dwarf MSPs.  Modelling fixed masses for both
the neutron star and the white dwarf, he concluded that observed
inclinations are not randomly distributed but rather tend to be high.
To explain this he suggested that pulsar beams preferentially lie parallel to
the orbital plane and so perpendicular to the pulsar spin axis.  The
following year \cite{1999Thorsett} used the $M_{\rm c} - P_{\rm orb}$
relation derived by \cite{1995Rappa} to conclude that the observed
inclinations of binary MSPs are consistent with a random distribution.
\cite{2005Stairs} similarly investigated a larger
data set with the $M_{\rm c} - P_{\rm orb}$ relations of both
\cite{theo} and \cite{1995Rappa}.  They concluded that the existing
forms of the $M_{\rm c} - P_{\rm orb}$ relations overestimate the
white dwarf masses at large periods and are also in conflict for the
shorter period system.  With a Kolmogorov--Smirnov (KS) test on the
cumulative distribution of orbital inclinations they found that the
$M_{\rm c}$--$P_{\rm orb}$ relation of \cite{theo} is incompatible at
the 99.5\% level when they drew the pulsar masses from a Gaussian centred on
$1.35\,\rm M_{\odot}$ with width $0.04\,\rm M_{\odot}$
\citep[mirroring the neutron star mass distribution found
  by][]{1999Thorsett}.  Repeating the same investigation for neutron
stars picked from a Gaussian centred on $1.75\,\rm M_{\odot}$ with
width $0.04\,\rm M_{\odot}$ they reached agreement at the 50\% level.
Our analysis here is
equivalent to that of \citet{2005Stairs} applied to a larger data set.

If the standard
recycling model is correct we expect the systems to be oriented
randomly in space so that the probability $P(i)\,{\rm d}i$ of finding an
inclination $i$ between $i$ and $i + di $ should be such that $P(i)
\propto \sin i$.  We are interested exclusively in the systems that
have periods within range of the $M_{\rm c}$--$P_{\rm orb}$
relations consistent with Case~B RLOF.  In the ATNF catalogue
the minimum possible mass $M_{\rm m}$, given the observed mass
function and a pulsar mass of $1.35\,\rm M_{\odot}$ is listed.
This can be transformed back to the binary mass function $f$ by
\begin{equation}
f =\frac{\left(M_{\rm d} \sin i \right)^{3}}{(M_{\rm a} +M_{\rm
    d})^{2}} = \frac{M_{\rm m} ^{3}}{(1.35\,{\rm M_\odot} + M_{\rm m})^{2}}, 
\end{equation}
where $M_{\rm a}$ is the actual mass of the pulsar.  We list the mass
functions $f$ in Table~\ref{mspsfull}, along with the companion masses
$M_{\rm d}$ obtained from $P_{\rm orb}$ with our canonical relation by
a spline interpolation.  For the combination of $f$ and $M_{\rm d}$
there is a maximum mass $M_{\rm a_{max}}$ of pulsar that can be
accommodated in the system.  We list in Table~\ref{mspsfull} too.
When $M_{\rm a_{max}} < 1.2\rm M_\odot$ we exclude the system from the
analysis.  Such cases are marked with bullets.  Note that all systems
found to have an ATNF minimum companion mass $M_{\rm d} > 0.5\rm
M_\odot$, too large to be a He white dwarf and marked with stars in
Table~\ref{mspsfull}, are automatically excluded by this cut.
Inclinations are then given by
\begin{equation}
\label{eqinc}
i = \sin^{-1}\left( \frac{(f(M_{\rm a} + M_{\rm d})^{2})^{1/3}}{M_{\rm d}} \right).
\end{equation}
and we expect $\cos i$ to be uniformly
distributed between 0~and~1 if the inclinations are consistent with
being picked from a uniform distribution.

\begin{figure}
\centering 
\includegraphics[width=8cm]{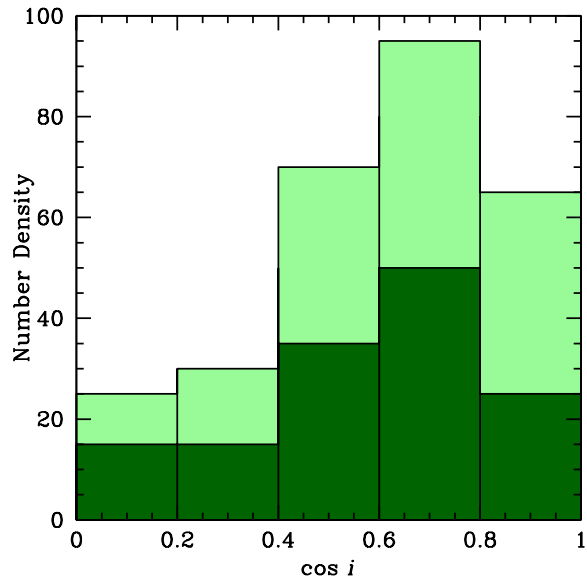} 
\caption{Histogram of the number density of the millisecond
pulsar binary systems with respect to $\cos i$, assuming a current neutron star
of mass $1.35\rm M_\odot$.  The solid histogram is just the low-period
systems (1 to~12\,d), the lightly shaded histogram adds all the remaining high-period systems.}
\label{hist135} 
\end{figure}

\begin{figure}
\centering 
\includegraphics[width=8cm]{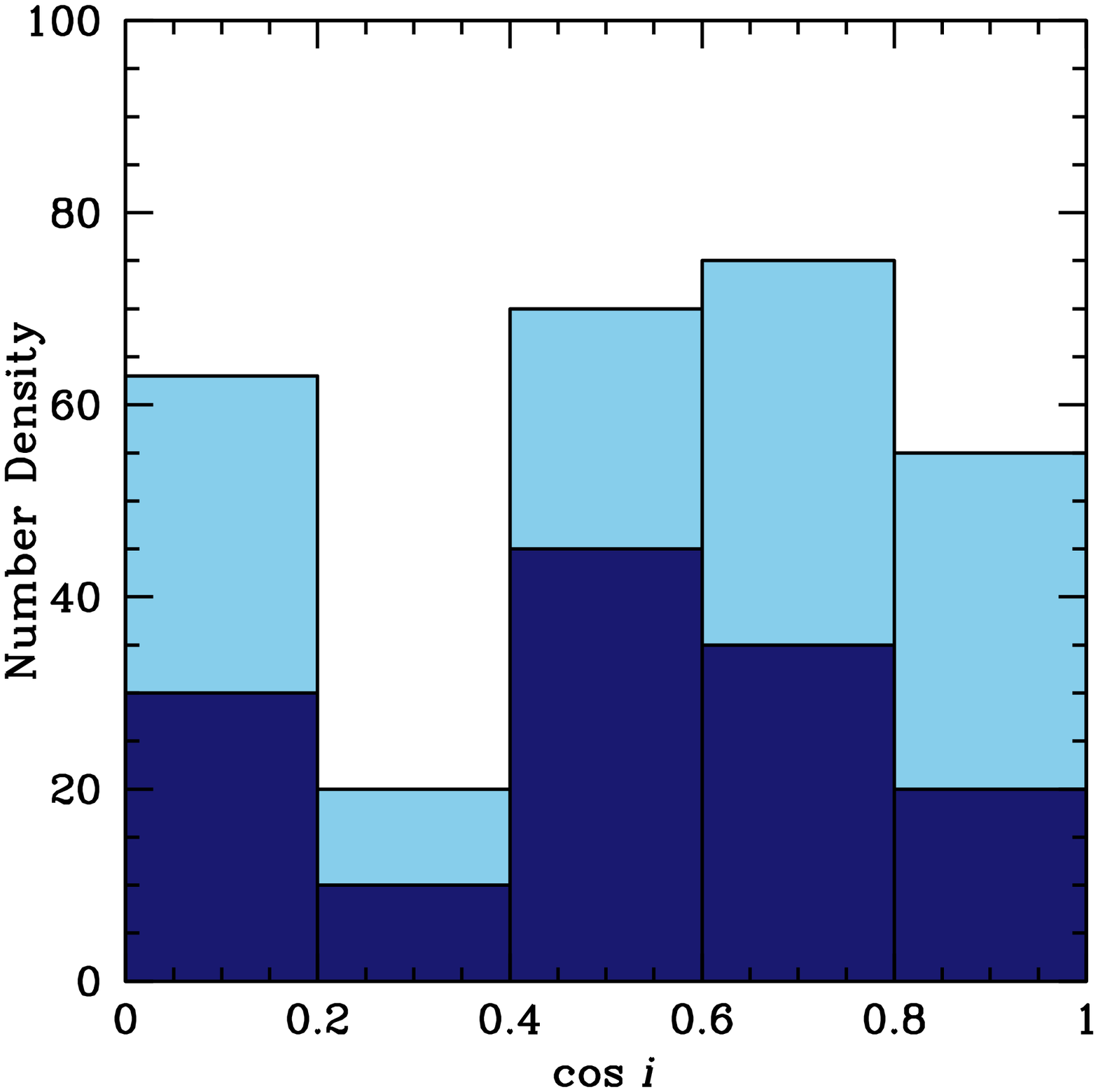} 
\caption{Histogram of the number density of the millisecond
pulsar binary systems with respect to $\cos i$, assuming a current neutron star
of mass $1.55\rm M_\odot$.  The solid histogram is just the low-period
systems (1 to~12\,d) and
the lightly shaded histogram adds all the remaining high-period systems.}
\label{hist155} 
\end{figure}

\begin{figure}
\centering
\includegraphics[width=8cm]{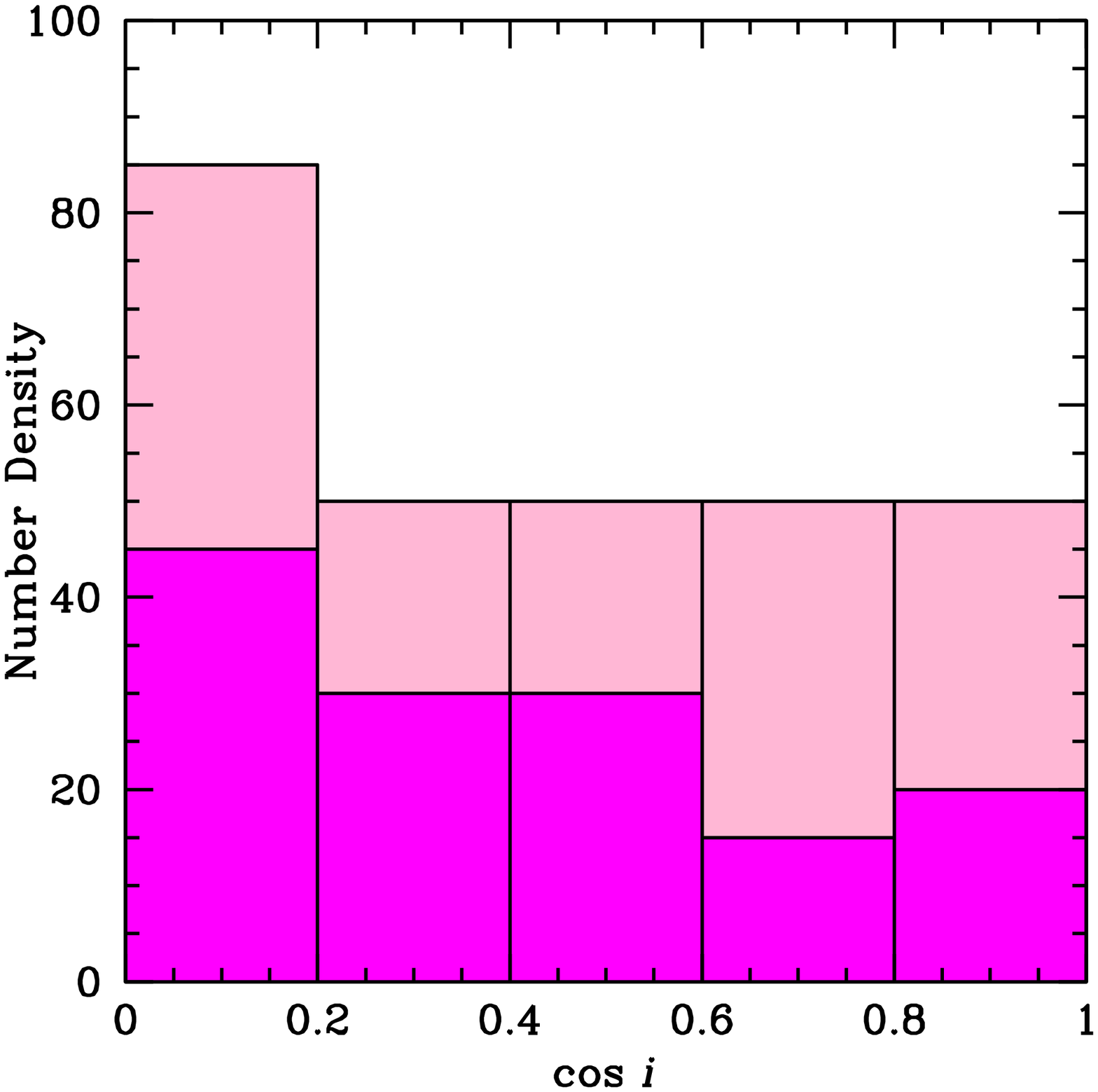} 
\caption{Histogram of the number density of the millisecond
pulsar binary systems with respect to $\cos i$, assuming a current neutron star
of mass $1.75\rm M_\odot$.  The solid histogram is just the low-period
systems (1 to~12\,d) and
the lightly shaded histogram adds all the remaining high-period systems.}
\label{hist175} 
\end{figure}

For each neutron star mass $M_{\rm a} \in \{1.35, 1.55, 1.75\}\rm
M_\odot$ we calculate the inclinations for all remaining systems in
Table~\ref{mspsfull}.  When the maximum permitted pulsar mass is
less than the chosen $M_{\rm a}$ we assign $i = 90^\circ$ and thereby
use $M_{\rm a} = M_{\rm a_{max}}$ for such systems.
Figs~\ref{hist135}, \ref{hist155} and~\ref{hist175} are histograms of
the numbers of BMSPs over $\cos i$ split into two orbital
period
ranges for the three different choices of neutron star mass.  There or
28~systems with $P_{\rm orb} < 12\,$d and 29~systems with longer
periods.  The
distributions are reasonably uniform but show a distinct dearth of
high inclination systems for $M_{\rm a} = 1.35\,\rm M_\odot$.  This
is less extreme as $M_{\rm a}$ is raised to $1.55\,\rm M_\odot$
and becomes an excess when $M_{\rm a} = 1.75\,\rm M_\odot$.  To
examine the significance of these trends we use Kolmogorov-Smirnov
tests on the cumulative distributions of $\cos i$.

\subsection{Kolmogorov--Smirnoff tests}

\begin{figure}
\centering 
\includegraphics[width=8cm]{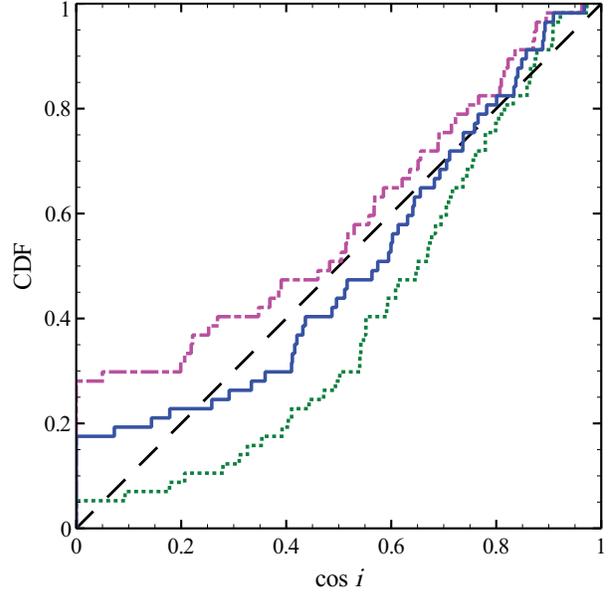} 
\caption{Cumulative distributions of $\cos i$.  The expected uniform
distribution is the dashed line.  The blue solid line is for the data when
we assume the pulsar masses are $1.55\,\rm M_{\odot}$ whenever
possible and as large as possible otherwise. 
The dotted line is for $1.35\,\rm M_{\odot}$ pulsars and 
the dot-dashed line for $1.75\,\rm M_{\odot}$ pulsars.
The up turn at $\cos i = 0$ is artificial because any system that
cannot accommodate a neutron star of the appropriate mass is placed
there.  This artificial difference is not used for our KS tests.} 
\label{figcumulative} 
\end{figure}

A KS test compares an observed cumulative distribution with a model.
In this case we compare the observed distribution of $\cos i$ with a
uniform distribution.
The KS statistic $D$ is the maximum difference between the observed
cumulative distribution and the uniform cumulative distribution for
any $\cos i$ and with this is associated the probability that $D$ of
this size or larger would be found for a sample of observations chosen
at random from the uniform distribution \citep{numrec}.
Fig.~\ref{figcumulative} shows the cumulative distributions of $\cos i$
for the three pulsar masses compared with the uniform distribution.
Recall that all the systems for which $M_{\rm a} > M_{\rm a_{max}}$
have been assigned $\cos i = 0$ and this raises the $\cos i = 0$ end of each
distribution in a perhaps biased manner.  The results for $M_{\rm a} =
1.35\,\rm M_\odot$ are however unaffected because the largest difference $D$
is at larger $\cos i$.  For the other two cases we take $D$
to be the largest difference beyond the first data point that has
$\cos i > 0$.  This eliminates the bias by treating the highest
inclination systems as if they were as uniformly distributed as
possible.

Tables~\ref{kstable135},~\ref{kstable155} and~\ref{kstable175} list
the $D$ and its significance for the three cases for all systems
combined and split into orbital period ranges.  We discard the system
in the middle with $P_{\rm orb} = 12.33\,$d so that the two ranges
have precisely the same number of systems and it is meaningful to
compare the KS~probabilities.  For $M_{\rm a} = 1.35\,\rm M_\odot$
there is a significant lack of high inclination systems and for
$M_{\rm a} = 1.75\,\rm M_\odot$ there is a significant excess and both
of these can be ruled out given their very low KS probabilities.
For $M_{\rm a} = 1.55\,\rm M_\odot$ a distribution as different as
observed would be found
35~times in 100~random samples so that the inclinations are consistent with
a uniform distribution of orientations in space.  
Equivalent analysis for $M_{\rm a} = 1.45\,\rm M_\odot$ gives a KS
probability of~0.040 while for $M_{\rm a} = 1.65\,\rm M_\odot$ it
is~0.173.  Neither of these are small enough to really rule out such
neutron star masses but we conclude that anything in the range around
$1.55 \le M_{\rm a}/{\rm M}_\odot \le 1.65$ is quite acceptable. 
The KS test is weak for small data sets
but this conclusion remains apparent for the low and high systems
taken independently.  When $M_{\rm a} = 1.55\,\rm M_\odot$ the high
probabilities of around one and two thirds mean there is no reason to
suggest that the distribution or orbital inclinations changes with
period in this case.  However when $M_{\rm a} = 1.75\,\rm M_\odot$ the
fit for the low-period group is somewhat better than for the
high-period group.  This is consistent with what \citet{2005Stairs}
found.  However the numbers are still too small for this to be a significant
conclusion, while the full distribution, likely to occur at random only
once in about 2,000~samples, does significantly rule out such a large neutron
star mass.

\begin{table}
\centering
\caption{KS test on $\cos i$ for $M_{\rm a} = 1.35\,\rm M_\odot$}
\begin{tabular}{|| l | c | c | c | r ||}
\hline
$M_{\rm a}/{\rm M_\odot}$ & $P_{\rm orb}/\rm d$ range & $D$ &
Probability& $N_{\rm sys}$  \\
\hline
1.35 & full & 0.242 & $0.000201$ & 57\\
1.35 & $< 12.3$ & 0.257 & $0.0406$ &28\\
1.35 & $> 12.4$ & 0.266 & $0.0307$ &28\\
\hline  
\label{kstable135}
\end{tabular}
\end{table}

\begin{table}
\centering
\caption{KS test on $\cos i$ for $M_{\rm a} = 1.55\,\rm M_\odot$}
\begin{tabular}{|| l | c | c | c | r ||}
\hline
$M_{\rm a}/{\rm M_\odot}$ & $P_{\rm orb}/\rm d$ range & $D$ & Probability& $N_{\rm sys}$  \\
\hline
1.55 & full & 0.120 & $0.359$ & 57\\
1.55 & $< 12.3$ & 0.181 & $0.286$ &28\\
1.55 & $> 12.4$ & 0.142 & $0.591$ &28\\
\hline  
\label{kstable155}
\end{tabular}
\end{table}

\begin{table}
\centering
\caption{KS test on $\cos i$ for $M_{\rm a} = 1.75\,\rm M_\odot$}
\begin{tabular}{|| l | c | c | c | r ||}
\hline
$M_{\rm a}/{\rm M_\odot}$ & $P_{\rm orb}/\rm d$ range & $D$ & Probability& $N_{\rm sys}$  \\
\hline
1.75 & full & 0.266 & $0.000470$ & 57\\
1.75 & $< 12.3$ & 0.122 & $0.772$ &28\\
1.75 & $> 12.4$ & 0.236 & $0.0747$ &28\\
\hline  
\label{kstable175}
\end{tabular}
\end{table}

\begin{figure}
\centering 
\includegraphics[width=8cm]{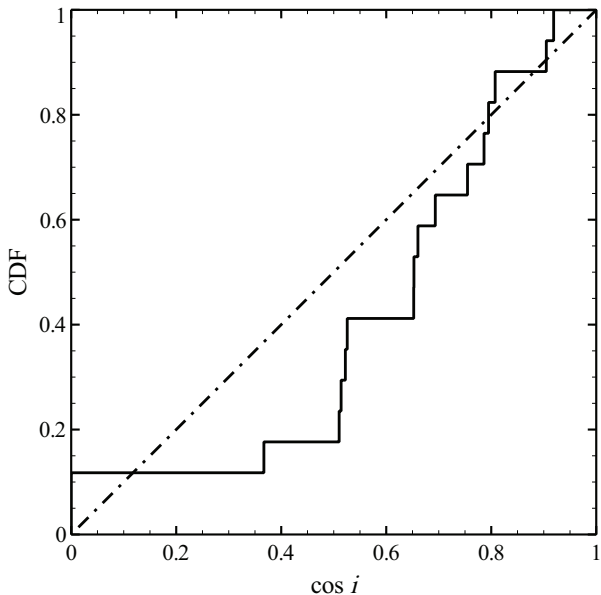} 
\caption{Cumulative distribution of $\cos i$ for the sample of 15~MSPs
  used by \citet{1998Backer} that fit our selection criteria.  The
  dot-dashed line is the CDF of the expected uniform distribution.
  The solid line is for Backer's sample of systems with $P_{\rm orb} >
  1\,$d obtained with our canonical $P_{\rm orb}-M_{\rm c}$ relation
  and a neutron star mass of $1.4\,\rm M_\odot$.
\label{backer}} 
\end{figure}

\begin{figure}
\centering
\includegraphics[width=8cm]{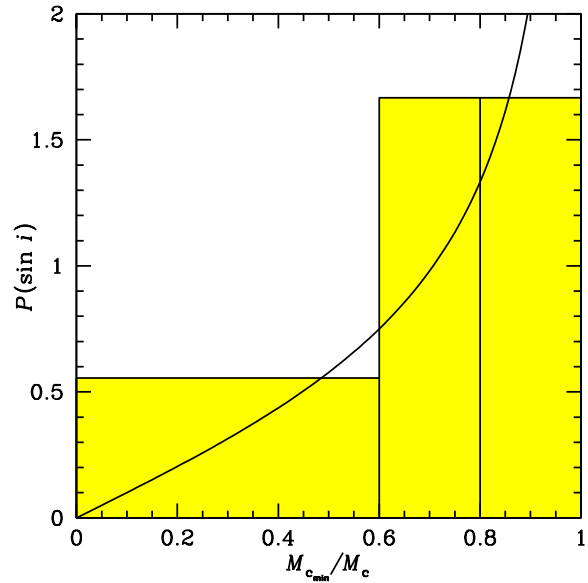}
\caption{Histogram of the ratio of the minimum companion mass to a
  pulsar of $1.4\,\rm M_\odot$ to its actual mass if it lies on our
  $P_{\rm orb}-M_{\rm c}$ relation compared to the distribution of
  $M_{\rm c_{\rm min}}/M_{\rm c}=\sin i$ for randomly oriented orbits (solid line).
\label{backersini}} 
\end{figure}

For comparison we apply the same analysis to the sample of MSPs used by
\cite{1998Backer} and with $M_{\rm a} = 1.4\,\rm M_\odot$, as he assumed.  Fig.~\ref{backer} shows the cumulative
distributions of $\cos i$ modelled with our canonical $P_{\rm
  orb}-M_{\rm c}$ and a neutron star mass of $1.4\,\rm M_\odot$, as
chosen by \citet{1998Backer}.  We exclude systems according to the
same criteria we have used for the full sample and assign inclinations
of $i = 90^\circ$ to systems with $M_{\rm a_{max}} < 1.4\,\rm
M_\odot$.  The KS probability of obtaining $D=0.334$ or larger is $3.44\times10^{-2}$
so that the data used by
\cite{1998Backer} actually show dearth of high inclinations just as we
find with $M_{\rm a} = 1.35\,\rm M_\odot$.  \citet{1998Backer} based
his conclusion on a comparison between the distribution of minimum
masses and a uniform distribution of $\sin i$.  However it is the
ratio of that minimum mass to the actual mass of the companion that
is $\sin i$.  In Fig.~\ref{backersini} we show the distribution of
this ratio when we assume that the actual companion mass is known from
$P_{\rm orb}$.  This distribution then shows a dearth of systems at
large $\sin i$ quite contrary to his original claim.
The conclusion that pulsar beams tend to
be perpendicular to their spin axis is therefore no longer tenable.

\section{Conclusion} 

We have demonstrated that the relation between orbital period and
companion white dwarf mass of recycled MSPs that have evolved via
Case~B RLOF is largely independent of initial donor mass, neutron star
mass and how conservative is the mass transfer.  It does however
depend on the metallicity of the donor star because this changes the
radius to core mass relation for red giants.  For a companion of
$1\,\rm M_\odot$, about the smallest that can be expected to evolve
within the Galactic field, the lowest orbital period BMSPs
formed without additional angular momentum loss have $P_{\rm orb}\approx
6.2\,$d.  However with magnetic braking, or some alternative angular
momentum loss mechanism, the relation can be extended down to $P_{\rm orb} <
1\,$d.

We selected all Galactic BMSPs, with $P_{\rm spin} < 30\,$ms, that
could have been recycled by accretion from a red giant with a helium
core that is now a He white dwarf companion from those listed in the
ATNF catalogue.  With our $P_{\rm orb}-M_{\rm c}$ relation we
calculated the orbital inclinations of these systems, assuming a set
of pulsar masses.  We find that when the pulsar mass is taken to be as
large as possible up to a maximum of $1.55\,\rm M_\odot$ the
distribution of inclinations is consistent with random orientation of
the orbits in space.  If the maximum pulsar mass is reduced to
$1.35\,\rm M_{\odot}$ there appears a dearth of high inclination systems
and if it is raised to $1.75\,\rm M_\odot$ there appears an excess.
Hence we deduce that all systems selected, and listed in
Table~\ref{mspsfull} without bullets, are consistent with having
pulsars of mass around $1.6\,\rm M_\odot$, having been recycled by
accretion from a red giant with a helium core and having no
observational bias towards particular orbital orientation.  Given
that selection of particular orbital inclinations would be encouraged
if MSP magnetic axes and hence their beams were preferentially aligned
with or orthogonal to their spin axes, our result is consistent with
random orientation of magnetic axes too.  Numbers of systems remain
too small to make conclusions about systems in different orbital
period ranges.

\section*{Acknowledgements}
SLS thanks STFC for her studentship. CAT thanks Churchill College for
his fellowship. LF and DTW thank the Institute of Astronomy for
supporting visits.  SLS thanks Philip Hall for help in making detailed
stellar models with mass transfer.  We also thank the referee for many
suggestions that have led to significant improvements to this work.

\appendix
\section{List of Selected BMSPs} 
Table~\ref{mspsfull} lists all ATNF millisecond pulsars, with $P_{\rm spin}
< 30\,$ms and $P_{\rm orb} > 1\,$d in the Galactic field.  
The fourth column lists the companion mass derived from our $P_{\rm
  orb}-M_{\rm c}$ relation.  The fifth column is the measured mass
function, reconstructed from the minimum mass given in the ATNF
catalogue.  The sixth column is the maximum neutron star mass that
could be accommodated with the given mass function and derived
companion mass.  Systems in which $M_{\rm a_{max}} < 1.2\,\rm M_\odot$
are marked and excluded from further analysis.  The final column
indicates whether there is any determination of the companion's type.

\begin{table*}
\centering
\caption{Binary millisecond pulsars in the ATNF Pulsar Catalogue in
  the Galactic field with $P_{\rm orb} > 1\,$d.  Systems with superscript $^{\rm B}$ were in the
  sample of \citet{1998Backer} and those
  with a subscript $_{\rm C}$ have measured masses (see Table~\ref{pulsartable1}).  The
  $^{\star}$ symbols denote systems with minimum masses in the ATNF
  catalogue larger than $0.5\,\rm M_{\odot}$.  The pulsar spin period
  is $P_{\rm spin}$ and $P_{\rm orb}$ is the orbital period.  The
  companion mass $M_{\rm c}$ is computed from $P_{\rm orb}$ with our
  canonical relation.  The measured mass function $f$ is calculated
  from $M_{\rm m}$ given in the ATNF catalogue and $M_{\rm
    a_{max}}$  is then the maximum pulsar mass that can be
  accommodated given $f$ and $M_{\rm c}$.  The $^{\bullet}$
  symbols denote systems for which
  $M_{\rm a_{max}} < 1.20\,\rm M_{\odot}$ and therefore do not fit the
    model.  The final column lists any information on or suggestion of
    the type of the companion to be found in references given in the
    ATNF catalogue.
MS = Main-sequence star, 
CO = CO or ONeMg white dwarf, He = He white dwarf, UL = Ultra-light companion 
or planet ($M_{\rm c} < 0.08\,\rm M_{\odot}$) and U = undetermined companion type. A letter T at the end 
denotes that there are more than two components in the system.  We excluded the systems with UL companions, 
MS companions and multiple systems.  Many of the MSPs with thought to have CO companions have minimum masses in the ATNF 
catalogue larger than $0.5\,\rm M_{\odot}$.  All are excluded because they require a minimum neutron 
star mass of less than $1.20\,\rm M_{\odot}$.
  \label{mspsfull}}
\begin{tabular}{|| l | l | l | l | l | l | l ||}
\hline
Name & $P_{\rm spin}/\rm ms$ & $P_{\rm orb}/\rm d$ & $M_{\rm c}/\rm
M_{\odot}$ & $f/\rm M_\odot$ & $M_{\rm a_{\rm max}}/\rm M_{\odot}$ & Comp. Type \\
\hline          
J0613--0200$^{\rm B}$        &       3.06184408653189  &              1.1985125753  &      0.187 &    0.00097185502& 2.408             &He\\
J1435--6100$^{\star}$        &       9.347972210248    &              1.3548852170  &      0.191 &    0.13832189   & 0.033$^{\bullet}$&CO\\
J2043+1711                  &       2.37987896026     &              1.482290809   &      0.194 &    0.0020928803 & 1.675             &He\\
J1909--3744$_{\rm C}$        &       2.9471080681076401&              1.533449474590&      0.195 &    0.0031219739 & 1.350            &He\\
J0034--0534$^{\rm B}$        &       1.8771818845850   &              1.589281801   &      0.197 &    0.0012634271 & 2.257         &He\\
J1622--6617                  &\!\!\!23.62344473927     &              1.640635183   &      0.198 &    0.00037472931& 4.346             &U\\
J0101--6422                  &       2.5731519721683   &              1.787596706   &      0.201 &    0.0016538449 & 2.008             &He\\
J1231--1411                  &       3.683878711077    &              1.860143882   &      0.202 &    0.0026445979 & 1.559            &He\\
J1949+3106$^{\star}$        &\!\!\!13.1381833437040   &              1.949535      &      0.203 &    0.10938610   & 0.073$^{\bullet}$&CO\\
J0218+4232$^{\rm B}$        &       2.3230904678309   &              2.028846084   &      0.204 &    0.0020384352 & 1.832             &He\\
J1439--5501$^{\star}$        &\!\!\!28.634888190455    &              2.117942520   &      0.205 &    0.22759446   &\!\!\!--0.011$^{\bullet}$&CO\\
J2017+0603                  &       2.896215815562    &              2.198481129   &      0.205 &    0.0023426589 & 1.716             &He\\
J2317+1439$^{\rm B}$        &       3.4452510710225   &              2.459331464   &      0.208 &    0.0021994253 & 1.808            &He\\
J1502--6752                  &\!\!\!26.7444237673      &              2.4844570     &      0.208 &    5.5708571e--06&\!\!\!39.908                &UL\\
B1802--07                    &\!\!\!23.10085528430     &              2.61676335    &      0.209 &    0.0094491902 & 0.774$^{\bullet}$&U\\
J1911--1114$^{\rm B}$        &       3.6257455713977   &              2.71655761    &      0.210 &    0.00079709848& 3.196             &He\\
J1431--5740                  &       4.1105439567658   &              2.726855823   &      0.210 &    0.0016887826 & 2.131             &U\\
J1748--3009                  &       9.684273          &              2.9338198     &      0.212 &    0.00028695598& 5.545             &He\\
J1216--6410                  &       3.539375658423    &              4.03672718    &      0.220 &    0.0016694467 & 2.300             &He\\
J1045--4509$^{\rm B}$        &       7.47422422621133  &              4.0835292547  &      0.220 &    0.0017649364 & 2.235             &He\\
J1337--6423$^{\star}$        &       9.423406713       &              4.78533407    &      0.224 &    0.10508201   & 0.103$^{\bullet}$&CO\\
J1745--0952                  &\!\!\!19.3763034411294   &              4.943453386   &      0.225 &    0.00059126344& 4.170             &He\\
J1732--5049                  &       5.31255028907845  &              5.262997206   &      0.226 &    0.0024490954 & 1.951            &He\\
J0721--2038                  &\!\!\!15.54239497400     &              5.46083280    &      0.227 &    0.029704982  & 0.400$^{\bullet}$&U\\
J0437--4715$^{\rm B}_{\rm C}$&       5.757451924362137 &              5.74104646    &      0.227 &    0.0012431183 & 2.835             &He\\
\hline
J1545--4550                  &       3.57528861884712  &              6.203064928   &      0.227 &    0.0015885663 & 2.480           &U\\
J1811--2405                  &       2.66059331690     &              6.27230204    &      0.227 &    0.0050692659 & 1.289             &He\\
J1603--7202$^{\rm B}$        &\!\!\!14.84195224908935  &              6.3086296703  &      0.227 &    0.0087882681 & 0.925$^{\bullet}$&U\\
J1017--7156                  &       2.3385144401138   &              6.5118988121  &      0.227 &    0.0028531163 & 1.800             &He\\
J2129--5721$^{\rm B}$        &       3.72634848296641  &              6.625493093   &      0.228 &    0.0010492057 & 3.123             &He\\
J1835--0114                  &       5.116387644239    &              6.6925427     &      0.228 &    0.0024171121 & 1.983             &He\\
J2145--0750                  &\!\!\!16.05242391433660  &              6.83893       &      0.228 &    0.024105367  & 0.474$^{\bullet}$&CO\\
J1022+1001$^{\star}$        &\!\!\!16.45292995078405  &              7.8051302826  &      0.233 &    0.083054673  & 0.157$^{\bullet}$&CO\\
J1543--5149                  &       2.05696039242     &              8.06077304    &      0.234 &    0.0044968769 & 1.453             &He\\
J0621+1002                  &\!\!\!28.853860730049    &              8.3186813     &      0.235 &    0.027026827  & 0.457$^{\bullet}$&CO\\
J1327--0755                  &       2.6779231971205   &              8.439086019   &      0.235 &    0.0044251757 & 1.479            &He\\
J1614--2230$_{\rm C}$       &       3.1508076534271   &              8.6866194196  &      0.236 &    0.020483367  & 0.565$^{\bullet}$&CO\\
J1125--6014                  &       2.6303807397848   &              8.75260353    &      0.236 &    0.0081279214 & 1.036$^{\bullet}$ &He\\
J1400--1438                  &       3.0842332007      &              9.54743       &      0.238 &    0.0070342534 & 1.149$^{\bullet}$ &He\\
J1841+0130                  &\!\!\!29.7727753332      &\!\!\!       10.471626      &      0.241 &    0.00042129066& 5.523             &He\\
J1918--0642                  &       7.64587288390884  &\!\!\!       10.9131777486  &      0.242 &    0.0052494287 & 1.403             &He\\
J1804--2717$^{\rm B}$        &       9.343030685703    &\!\!\!       11.1287115     &      0.243 &    0.0033469189 & 1.825             &He\\
B1855+09$^{\rm B}_{\rm C}$  &       5.362             &\!\!\!       12.32719       &      0.246 &    0.0055573963 & 1.389            &He\\
J1900+0308                  &       4.909239016418    &\!\!\!       12.47602144    &      0.246 &    0.0020899584 & 2.425            &He\\
J2236--5527                  &       6.907549392921    &\!\!\!       12.68918715    &      0.247 &    0.0045070041 & 1.578            &U\\
J1933--6211                  &       3.543431438847    &\!\!\!       12.81940650    &      0.247 &    0.012103419  & 0.869$^{\bullet}$&U\\
J1600--3053                  &       3.59792850865547  &\!\!\!       14.3484577709  &      0.250 &    0.0035560324 & 1.851             &He\\
J1741+1351                  &       3.7471544         &\!\!\!       16.335         &      0.254 &    0.0053997280 & 1.491             &He\\
J0900--3144                  &\!\!\!11.1096491573889   &\!\!\!       18.73763576    &      0.259 &    0.015693887  & 0.795$^{\bullet}$&U\\
J1801--3210                  &       7.45358437341     &\!\!\!       20.7716995     &      0.263 &    0.0011851566 & 3.646             &He\\
J1709+2313                  &       4.6311962778409   &\!\!\!       22.71189238    &      0.265 &    0.0074382973 & 1.319             &He\\
J1618--39                    &\!\!\!11.987313          &\!\!\!       22.8           &      0.265 &    0.0022177665 & 2.638             &He\\
B1257+12                    &       6.21853194840048  &\!\!\!       25.262         &      0.26856700 &    0.0000000    & $\infty$              &UL\\
\end{tabular}
\end{table*}

\begin{table*}
\centering
\contcaption{The binary millisecond pulsars in the ATNF Pulsar Catalogue that are in the galactic field.}
\begin{tabular}{|| l | l | l | l | l | l | l ||}
\hline
Name & $P_{\rm spin}/\rm ms$ & $P_{\rm orb}/\rm d$ & $M_{\rm c}/\rm
M_{\odot}$ & $f/\rm M_\odot$ & $M_{\rm a_{\rm max}}/\rm M_{\odot}$ & Comp. Type\\
\hline          
J1844+0115                  &       4.185543936664    &\!\!\!       50.6458881     &      0.295 &    0.0011918677 & 4.341             &He\\
J1825--0319                  &       4.553527919736    &\!\!\!       52.6304992     &      0.296 &    0.0023624627 & 3.020            &U\\
J0614--3329                  &       3.148669579439    &\!\!\!       53.5846127     &      0.297 &    0.0078951142 & 1.523             &He\\
J2033+1734                  &       5.9489575348502   &\!\!\!       56.30779527    &      0.299 &    0.0027759932 & 2.799             &He\\
J1910+1256$_{\rm C}$        &       4.98358394056511  &\!\!\!       58.466742029   &      0.300 &    0.0029628574 & 2.720             &He\\
J1713+0747$^{\rm B}_{\rm C}$&       4.570136525082782 &\!\!\!       67.8251298718  &      0.306 &    0.0078961829 & 1.595             &He\\
J1455--3330$^{\rm B}$        &       7.987204796261    &\!\!\!       76.1745676     &      0.310 &    0.0062715859 & 1.869            &He\\
J1125--5825                   &       3.1022139189504   &\!\!\!       76.4032169     &      0.310 &    0.0070010555 & 1.754          &He\\
J2019+2425$^{\rm B}$         &       3.93452408033124  &\!\!\!       76.51163479    &      0.310 &    0.010686460  & 1.361         &He\\
J1737--0811                   &       4.1750173128551   &\!\!\!       79.517379      &      0.312 &    0.00013803925&\!\!\!14.497       &U\\
J1850+0124                   &       3.559763768043    &\!\!\!       84.949858      &      0.314 &    0.0058483603 & 1.989          &He\\
J1935+1726                   &       4.200101791882    &\!\!\!       90.76389       &      0.317 &    0.0042604451 & 2.416          &He\\
J2229+2643$^{\rm B}$         &       2.9778192947556   &\!\!\!       93.0158934     &      0.318 &    0.00083949925& 5.866            &He\\
J1903+0327$^{\star}$         &       2.14991236434921  &\!\!\!       95.174118753   &      0.319 &    0.13955857   & 0.163$^{\bullet}$&MST\\
J1751--2857                   &       3.9148731963690   &\!\!\!\!\!\!110.7464576     &      0.325 &    0.0030130429 & 3.051           &He\\
J1853+1303$_{\rm C}$         &       4.09179738145616  &\!\!\!\!\!\!115.65378643    &      0.327 &    0.0054396584 & 2.207          &He\\
B1953+29$^{\rm B}$           &       6.1331665102401   &\!\!\!\!\!\!117.34909728    &      0.327 &    0.0024167741 & 3.485          &He\\
J2302+4442                   &       5.192324646411    &\!\!\!\!\!\!125.935292      &      0.331 &    0.0092095292 & 1.650             &U\\
J1643--1224$^{\rm B}$         &       4.62164151699818  &\!\!\!\!\!\!147.01739776    &      0.338 &    0.00078297226& 6.671            &He\\
J1708--3506                   &       4.50515894826     &\!\!\!\!\!\!149.13318       &      0.338 &    0.0018287070 & 4.261            &He\\
J1640+2224$^{\rm B}$          &       3.16331581791380  &\!\!\!\!\!\!175.46066194    &      0.346 &    0.0059074572 & 2.302            &He\\
B1620--26                     &\!\!\!11.0757509142025   &\!\!\!\!\!\!191.44281       &      0.350 &    0.0079748112 & 1.972             &HeT\\
J0407+1607                   &\!\!\!25.70173919463     &\!\!\!\!\!\!669.0704        &      0.437 &    0.0028931881 & 4.939             &He\\

\hline  
\end{tabular}
\end{table*}


\label{lastpage}
\end{document}